\documentclass{IEEEtran4PSCC}

\usepackage[pdftex]{graphicx}
\usepackage[cmex10]{amsmath}
\usepackage{shellesc}
\usepackage{cite}
\usepackage{amsmath,amssymb,amsfonts}
\usepackage{algorithmic}
\usepackage{graphicx}
\usepackage{textcomp}
\usepackage{xcolor}
\usepackage{mathtools}
\usepackage{siunitx}
\usepackage{tikz}
\usepackage{booktabs}     
\usepackage{array}       
\usepackage{caption}      
\usetikzlibrary{calc,shapes,arrows.meta,positioning}

\DeclareSIUnit{\pu}{pu}
\DeclareSIUnit{\voltampere}{VA}

\usepackage{booktabs}
\usepackage[nolist]{acronym}
\usepackage[switch]{lineno}

\usepackage{array}
\usepackage{multirow}
\newcolumntype{L}[1]{>{\raggedright\let\newline\\\arraybackslash\hspace{0pt}}m{#1}}

\makeatletter
\let\old@ps@headings\ps@headings
\let\old@ps@IEEEtitlepagestyle\ps@IEEEtitlepagestyle
\def\psccfooter#1{%
    \def\ps@headings{%
        \old@ps@headings%
        \def\@oddfoot{\strut\hfill#1\hfill\strut}%
        \def\@evenfoot{\strut\hfill#1\hfill\strut}%
    }%
    \def\ps@IEEEtitlepagestyle{%
        \old@ps@IEEEtitlepagestyle%
        \def\@oddfoot{\strut\hfill#1\hfill\strut}%
        \def\@evenfoot{\strut\hfill#1\hfill\strut}%
    }%
    \ps@headings%
}
\makeatother

\psccfooter{%
        \parbox{\textwidth}{\hrulefill \\ \small{24th Power Systems Computation Conference} \hfill \begin{minipage}{0.2\textwidth}\centering \vspace*{4pt} \includegraphics[scale=0.06]{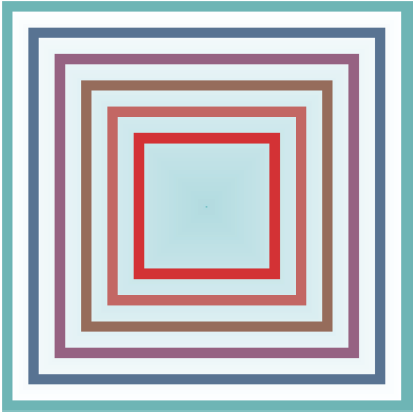}\\\small{PSCC 2026} \end{minipage} \hfill \small{Limassol, Cyprus --- June 8-12, 2026}}%
}

\begin{document}
%
\title{Feature Selection for Fault Prediction in Distribution Systems}

 \author{\IEEEauthorblockN{Georg Kordowich\IEEEauthorrefmark{1}, Julian Oelhaf \IEEEauthorrefmark{3}, Siming Bayer \IEEEauthorrefmark{3}, Andreas Maier \IEEEauthorrefmark{3}, Matthias Kereit\IEEEauthorrefmark{2}, and Johann Jaeger\IEEEauthorrefmark{1}}
	\IEEEauthorblockA{\IEEEauthorrefmark{1} Institute of Electrical Energy Systems, 
		Friedrich-Alexander-Universität Erlangen-Nürnberg, Erlangen, Germany}
	
	\IEEEauthorblockA{\IEEEauthorrefmark{3} Pattern Recognition Lab, Friedrich-Alexander-Universität Erlangen-Nürnberg, Erlangen, Germany}
	\IEEEauthorblockA{\IEEEauthorrefmark{2} Siemens AG, Berlin, Germany \\ georg.kordowich@fau.de}
}

\maketitle
\begin{abstract}
While conventional power system protection isolates faulty components only after a fault has occurred, fault prediction approaches try to detect faults before they can cause significant damage. Although initial studies have demonstrated successful proofs of concept, development is hindered by scarce field data and ineffective feature selection. To address these limitations, this paper proposes a surrogate task that uses simulation data for feature selection. This task exhibits a strong correlation (r = 0.92) with real-world fault prediction performance. We generate a large dataset containing 20000 simulations with 34 event classes and diverse grid configurations. From 1556 candidate features, we identify 374 optimal features. A case study on three substations demonstrates the effectiveness of the selected features, achieving an F1-score of 0.80 and outperforming baseline approaches that use frequency-domain and wavelet-based features.
\end{abstract}


\thanksto{\noindent \textcopyright 2026. This manuscript version is made available under the CC-BY-NC-ND 4.0 license https://creativecommons.org/licenses/by-nc-nd/4.0/ \\
Submitted to the 24th Power Systems Computation Conference (PSCC 2026).}

\begin{figure*}
	\centering
	\includegraphics[trim={0 0 0 0},clip,width=\linewidth]{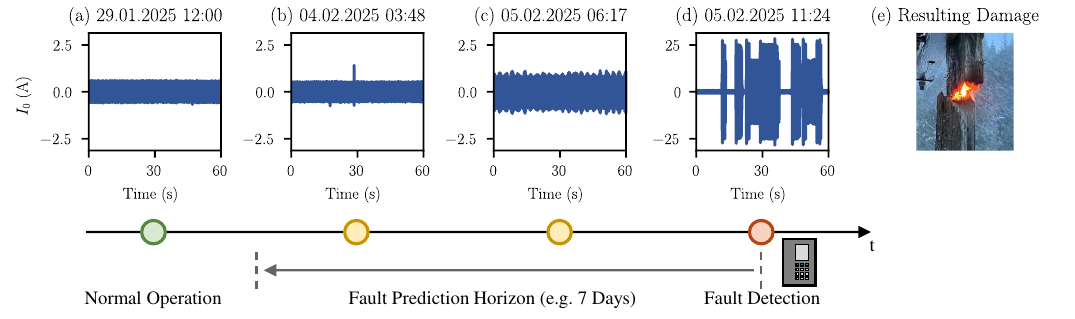}
	\caption{Zero-sequence current measurements from a \qty{20}{\kilo\volt} grid showing normal operation (a) and exemplary pre-fault precursor symptoms (b-d). The fault was caused by a conductor falling onto a wooden pole, which then ignited (e).}
	\label{fig:fig1}
\end{figure*}
\section{Introduction}
\subsection{Fault Prediction in Power Systems}
Conventional power system protection isolates faulty components only after a fault has occurred. This reactive approach is suboptimal when precursor symptoms could enable proactive measures before fault development causes significant destruction.
Generally, these grid faults can be categorized into two groups: (i) unsystematic (unpredictable) faults and (ii) systematic (predictable) faults~\cite{dashtiSurveyFaultPrediction2021}.

Unsystematic faults, caused by external influences such as excavators, animals, or lightning, are inherently unpredictable. In contrast, systematic faults result from developing physical phenomena and can be predicted by analyzing electrical measurements. Examples include contaminated insulators that eventually cause arc faults, degradation of cable insulation, or high-impedance faults caused by vegetation.

Fault prediction (FP) aims to detect systematic faults before they cause significant damage by identifying precursor symptoms in voltage and current waveforms. Due to the complexity of detecting subtle, varying precursor symptoms and the limited immediate impact of misclassifications, recent research has focused on machine‑learning approaches.

Fig.~\ref{fig:fig1} demonstrates the need for FP: In this case, a conductor fell from an insulator onto a wooden pole, initially causing low-current arcing that eventually led to ignition. Protection systems only detected this high-impedance fault after the pole was already burning, even though precursor symptoms are observable more than \qty{32}{\hour} earlier.

The prediction horizon shown in Fig.~\ref{fig:fig1} defines the temporal window during which a fault is considered "imminent". Effective FP systems should predict high fault probabilities within, and low probabilities outside this horizon. FP approaches are trained by utilizing historical protection relay operations, defining the label as \textit{True}, if measurements fall within the pre-fault prediction horizon and \textit{False}, otherwise.

The increasing data availability in smart grids facilitates these approaches~\cite{rubioSmartGridProtection2025}. This has prompted the \textit{IEEE PES Working Group on Power Quality and Data Analytics} to call for further research on FP in a recent report~\cite{powerqualitydataanalyticsworkinggroupElectricSignaturesPower2019}.

\subsection{Related Literature}
Recent review papers by Dashti et al. \cite{dashtiSurveyFaultPrediction2021}, Imam et al.~\cite{imamParametricNonparametricMachine2024}, and Haleem et al. \cite{haleemmedattilibrahimIncipientFaultDetection2024}, provide an overview of current trends in the literature. Five topics are loosely related to FP, namely partial discharge monitoring, predictive maintenance, power quality, event detection and incipient fault detection.

Although a wide range of sensors could be used for FP, the high investment cost of deploying additional sensors presents a significant challenge~\cite{haleemmedattilibrahimIncipientFaultDetection2024}. Therefore, we focus on utilizing existing voltage and current measurements available via IEC 61850 on the process bus. \textit{Partial discharge monitoring}, requiring additional high frequency sensors, is therefore not considered in this work.

One straightforward approach to prevent outages is to employ \textit{predictive maintenance} which uses low sample rate operational records to optimize long term maintenance schedules~\cite{alvarez-alvaradoPowerSystemReliability2022}. However, these approaches do not capture precursor symptoms in real-time measurements. FP is a more direct approach that exploits real‑time sensor data to predict faults based on precursor symptoms. Although these anomalies are related to \textit{power quality} phenomena, it is important to distinguish FP as a separate task, as precursor symptoms of faults are not necessarily power quality concerns~\cite{powerqualitydataanalyticsworkinggroupElectricSignaturesPower2019}.

\textit{Event Detection} approaches aim to detect anomalies in measurement data and subsequently try to classify the event type. Previous literature includes approaches both on simulation and real-world data and a wide range of faults and operating events\cite{wilsonGridEventSignature2024, alacaEventTypeIdentificationPower2024}. One focus is the detection of high-impedance faults, as they are particularly difficult to detect with conventional methods \cite{ehsaniConvolutionalAutoencoderAnomaly2022, baquiHighImpedanceFault2011, ledesmaTwolevelANNbasedMethod2020}. However, these approaches do not explicitly aim to relate these events to subsequent faults.

Therefore, \textit{Incipient fault detection} (IFD) is the most closely related field to FP as its aim is to detect and classify early-stage anomalies and also map them to subsequent faults~\cite{liIncipientFaultDetection2023}. By classifying these early-stage anomalies, recent papers demonstrated IFD in cables~\cite{mousaviNovelConditionAssessment2009, luCableIncipientFault2022}, overhead lines~\cite{dasilvaNewMethodologyMultiple2018,liIncipientFaultIdentification2021}, transformers~\cite{butler-purryIdentifyingTransformerIncipient2003, bhowmickOnlineDetectionInterturn2015}, or capacitors~\cite{heElectricSignatureDetection2021}. A disadvantage of IFD is its dependence on labeled training data: Each electric signature must be explicitly categorized into different classes (e.g., \textit{incipient fault}, \textit{capacitor switching}, \textit{inrush}, etc.)~\cite{luCableIncipientFault2022}. Therefore, experts must either create highly detailed simulations covering many possible fault and operating scenarios, or manually review and annotate large amounts of real-world field data.

\textit{Fault prediction} approaches avoid the need for manual labeling by utilizing historical protection relay operations to directly estimate the probability of future faults without requiring intermediate classification of electrical signatures. Zhang et al.~\cite{zhangDataBasedLineTrip2018} and Skydt et al.~\cite{skydtProbabilisticSequenceClassification2021} demonstrated the feasibility of such approaches on phasor-domain data. Li et al.~\cite{liDataDrivenFrameworkIncipient2025} later showed that point-on-wave measurements in combination with feature extraction are critical to improve the identification of precursor symptoms of faults. Balouji et al.~\cite{baloujiDistributionNetworkFault2023} also stressed the importance of waveform data and feature engineering. Chang and Li~\cite{changHybridIntelligentApproach2019,liMultipleAnomalyDetection2022} proposed a similar approach, that relies mainly on voltage waveforms. While these approaches cannot predict unsystematic, inherently unpredictable faults, previously mentioned papers demonstrate their value in preventing outages and improving grid resiliency by predicting systematic, developing faults.

\subsection{Contribution}
Despite significant advancements in the field of fault prediction, several challenges remain. The scarcity of labeled field data is an obstacle because extensive experiments on limited data risks overfitting~\cite{haleemmedattilibrahimIncipientFaultDetection2024}. Additionally, despite the critical importance of feature extraction, as of now a diverse set of features is utilized by different studies and no comprehensive comparison or feature selection has been made~\cite{imamParametricNonparametricMachine2024}.

The primary objective of this work is to advance FP in distribution systems. More specifically, we want to address the identified research gaps with the following contributions: 
\begin{itemize}
	\item[1)] Proposing and validating a surrogate task for FP on simulation data to avoid experiments on scarce field data.
	\item[2)] A comprehensive feature selection for fault prediction.
	\item[3)] A case study on real-world data from three substations to demonstrate the efficiency of the selected features.
\end{itemize}

\subsection{Paper Organization}
For this purpose, the paper is structured as follows: First, an overview of the FP pipeline is given in Sec.~\ref{sec:methods}. The creation of a simulation dataset we utilize for a surrogate task for feature selection is explained in Sec.~\ref{sec:simdata}. After the feature selection and its results are discussed in Sec.~\ref{sec:feat_selec}, a case study on the real-world dataset is shown in Sec.~\ref{sec:casestudy}. Finally, the results are discussed in Sec.~\ref{sec:discussion} and a conclusion is drawn in Sec.~\ref{sec:conclusion}.

\section{Overview of a Machine Learning Based Fault Prediction Pipeline}
\label{sec:methods}
The goal of FP is to improve grid reliability by providing early warnings of developing faults before they can cause significant damage. For this purpose, we utilize the setup shown in Fig.~\ref{fig:fig2}, an adapted version from~\cite{baloujiDistributionNetworkFault2023,changHybridIntelligentApproach2019,liMultipleAnomalyDetection2022}. The data‑processing pipeline for deployment and offline training is largely identical. The main difference is offline training requiring feature selection and classifier training, while online prediction involves feature extraction and classifier inference.
\subsection{Data Aggregation}
\label{sec:data_agg}
\begin{figure}
	\centering
	\includegraphics[trim={0 0 0 0},clip,width=\linewidth]{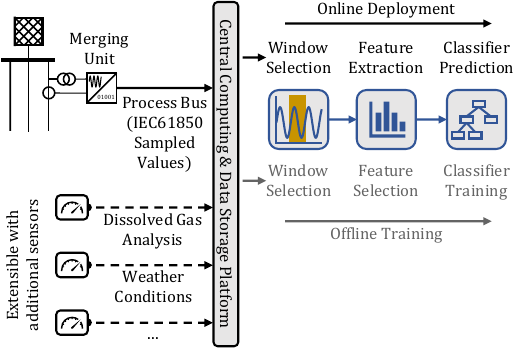}
	\caption{Fault prediction pipeline during training (gray) and deployment (black). Training is executed offline on recorded data, while deployment processes data in real-time. The pipeline utilizes voltage and current measurements could also incorporate additional data from different sources.}
	\label{fig:fig2}
\end{figure}
Voltage and current measurements are digitized using existing merging units and streamed on the process bus as sampled values as defined in IEC 61850-9-2. From the process bus, they are continuously recorded for online fault detection and simultaneously stored in a database for offline training. For this study, this setup was implemented at three substations in different \qty{22}{\kilo\volt} distribution grids resulting in approximately \qty{14}{\tera\byte} of \qty{14.4}{\kilo\hertz} measurements. Each discrete sample $\mathbf{m}$ in the dataset contains eight channels: three phase voltages, the zero-sequence voltage, three phase currents, and  the zero-sequence current:
\begin{align}
	\mathbf{m} = \Big[v_{\mathrm{a}}, v_{\mathrm{b}}, v_{\mathrm{c}}, v_{\mathrm{0}}, i_{\mathrm{a}}, i_{\mathrm{b}}, i_{\mathrm{c}}, i_{\mathrm{0}} \Big]
	\label{eq:time_vector}
\end{align}

The recorded measurements include diverse operating conditions and events as well as different fault types, for example three-phase, two-phase and single-line-to-ground faults.

This study focuses on FP based on voltage and current measurements, but Fig.~\ref{fig:fig2} shows that the pipeline can be extended using additional data sources from heterogenous sensors deployed in digital substations as described in~\cite{liAutonomousSmartGrid2023}.

\subsection{Window Selection}
As precursor symptoms of faults are typically intermittent, processing the entire data is both unnecessary and impractical. Therefore, prior approaches contain a pre-filtering step to differentiate between normal operation and anomalous windows~\cite{baloujiDistributionNetworkFault2023,changHybridIntelligentApproach2019,liMultipleAnomalyDetection2022}. However, such methods can require manual anomaly labeling for training~\cite{baloujiDistributionNetworkFault2023} and risk excluding informative precursor signals.

In contrast, we simply identify the two most relevant \qty{500}{\milli\second} windows per minute. While some of these windows will contain normal operation data that can slightly reduce the signal-to-noise ratio for the FP task, this approach eliminates manual labeling requirements and prevents accidental filtering of precursor symptoms over extended periods of time.

We employ two methods to detect both continuous changes and short transients. For continuous changes, we adapt  the approach of Oelhaf et al.~\cite{oelhafUnsupervisedClusteringFault2025} decomposing each channel into trend $T$, seasonal $S$, and residual $e$ components. The most relevant \qty{500}{\milli\second} window is then selected by identifying the window with the maximum root mean square (RMS) of the normalized trend component, where normalization uses the mean $\mu_{c}$ and standard deviation $\sigma_{c}$ of the trend $T$ computed over one minute of data per channel $c$:
\begin{align}
	X[t] &= T[t] + S[t] + e[t]\\
	\text{Window Continuous}
	&= \arg\max_{w_c \in \mathcal{W}}
	\left( \underset{t \in w_c}{\operatorname{RMS}}\!\left(
	\frac{T[t] - \mu_c}{\sigma_c}
	\right) \right)
	\label{eq:rms_window}
\end{align}

We detect transient peaks by selecting the window with the highest crest factor, adapting the threshold-based approach of Mousavi and Butler-Purry~\cite{mousaviNovelConditionAssessment2009}:
\begin{align} 
	\text{Window Transient} = \arg\max_{w_c \in \mathcal{W}} \left( \frac{ \max_{t \in w_c} |X[t]| }{ \operatorname{RMS}_{t \in w_c} (X[t]) } \right) 
	\label{eq:peak_window} 
\end{align}

\begin{figure}
	\centering
	\includegraphics[width=1\linewidth]{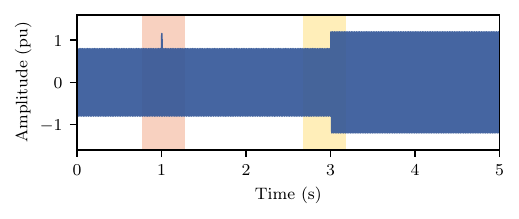}
	\caption{Identification of the most relevant windows for an exemplary signal, where the selected continuous window is marked in yellow and the transient window in red.}
	\label{fig:signalwindows}
\end{figure}

This results in two windows of size $(8 \times 7200)$, each containing eight channels with \qty{500}{\milli\second} of data. The window selection is visualized for an exemplary signal in Fig.~\ref{fig:signalwindows}.

\subsection{Feature Extraction}
Feature extraction has been identified as a critical step for FP in prior studies, making it the primary focus of this work. The main challenge for feature selection is the limited number of fault occurrences in real-world datasets, as extensive experiments directly on field data induces a high risk of overfitting. To address this, we perform feature selection experiments using a surrogate task on simulation data.

As a surrogate task we selected the classification of transient events in power systems. This is based on the hypothesis that features, which are useful to differentiate between different transient events are also useful to identify the subtle precursor symptoms of faults. Sec.~\ref{sec:feat_selec} details the feature selection methodology and validates this hypothesis, while Sec.~\ref{sec:simdata} describes the simulation framework and event scenarios.

Crucially, simulation data is used exclusively for feature selection. The final FP classifier is trained solely on real-world data. This is beneficial as it avoids the risk of domain shift when using simulation data for training but deploying the model on real-world data. Here, domain shift refers to the differences between the source domain (simulation data) and the target domain (real-world data), which can lead to a significant amount of misclassifications and therefore render the ML model inaccurate.

\subsection{Fault Prediction}
Following the feature extraction process described in Sec.~\ref{sec:feat_selec}, 374 features are extracted for each \qty{500}{\milli\second} window. To give the subsequent classifier temporal context which captures the progression of precursor symptoms over time, we aggregate these features hourly using the minimum, maximum, mean, and standard deviation. This aggregation yields 1496 features per hour (374 features × 4 aggregations).

The goal of FP is to estimate the probability of a fault occurring in the subsequent fault prediction horizon of a given window. Therefore, each hourly window is assigned a binary label $y_t \in \{0, 1\}$, where $y_t = 1$ indicates a fault occurs within seven days following the window, and $y_t = 0$ denotes no fault within this period. This seven-day prediction horizon is adopted from \cite{baloujiDistributionNetworkFault2023} and will be optimized in future work.

For FP, we use a random forest classifier (RF), which has demonstrated robust performance and resilience to noise for similar classification tasks~\cite{liDataDrivenFrameworkIncipient2025,oelhafImpactDataSparsity2025,sahooEnhancedIncipientFault2025}.

\section{Simulation Dataset}
\label{sec:simdata}
\begin{figure}
	\centering
	\includegraphics[trim={0 0 0 0},clip,width=0.95\linewidth]{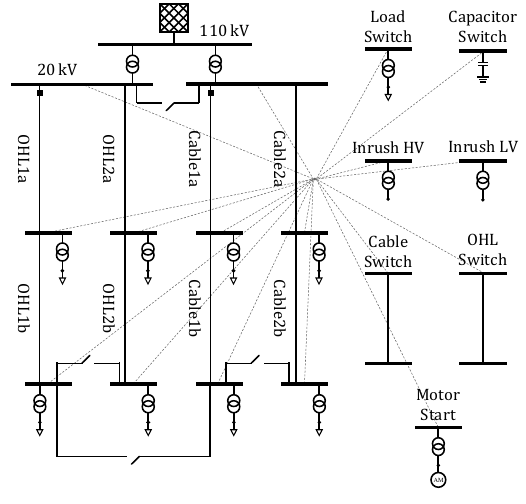}
	\caption{Distribution grid model used for simulation data generation. Event buses are dynamically connected to random network nodes during simulation.}
	\label{fig:power-system}
\end{figure}
To address the scarcity of labeled field data for feature selection, we utilize the classification of various transient events as a surrogate task. For the surrogate task to be effective, a wide variety of operating and fault scenarios must be included in the dataset to prevent overfitting on specific grids or events. Therefore, we utilize a custom simulation framework which is explained in detail in the following subsections.

\subsection{Grid Model Configuration}
The grid model, implemented in DIgSILENT's PowerFactory~\cite{digsilentPowerFactory2024}, is loosely based on the European configuration of the CIGRE medium-voltage distribution network benchmark system~\cite{BenchmarkSystemsNetwork2014}. As shown in Fig.~\ref{fig:power-system}, the \qty{22}{\kilo\volt} network consists of two overhead line (OHL) and two underground cable feeders, each supplying distributed loads. Switch states are randomized to generate diverse meshing configurations, increasing operational diversity. Geometric conductor models are applied to OHLs to accurately represent phase asymmetries that induce zero-sequence voltages during normal operation. Cables are represented by standard symmetric models. This configuration ensures realistic representation of both radial and weakly meshed distribution grid.

\subsection{Event Simulation}
As IFD is the most closely related topic, we identified the 17 discrete event types listed in Tab.~\ref{tab:simulated-events} from previous literature on IFD~\cite{luCableIncipientFault2022, liIncipientFaultIdentification2021, zhangMulticycleIncipientFault2017}. In each simulation one of the 17 events is selected. These events include all types of permanent faults with constant fault resistances up to \qty{20}{\ohm} (1PHG SHC, 2PH SHC, 2PHG SHC, 3PH SHC). Additionally, we simulate single-phase high-impedance faults with resistances up to \qty{150000}{\ohm} as well as incipient faults characterized by self extinguishing behavior and arc characteristics, which are represented using Kizilcay's dynamic arc model~\cite{kizilcayDigitalSimulationFault1991,zhangMulticycleIncipientFault2017}. Additionally, we simulate various operating events, which include energizing previously disconnected cables, overhead lines, capacitors, loads, motors and transformers to higher and lower voltage levels. Additionally, all respective elements can be switched off after they have previously been connected to the grid. Switching events are simulated at any of the \qty{20}{\kilo\volt} buses, while faults occur at random locations along overhead lines or cables.

\subsection{Parameter Randomization}
\begin{table}
	\centering
	\caption{List of Simulated Events}
	\label{tab:simulated-events}
	\begin{tabular}{%
			l 
			>{\raggedright\arraybackslash}p{0.65\linewidth} 
		}
		\toprule
		\textbf{Event name} & \textbf{Event description} \\ 
		\midrule
		Inrush LV Trafo & Inrush of a transformer connecting the \qty{20}{\kilo\volt} grid to a lower voltage level. \\
		Inrush HV Trafo & Inrush of a transformer connecting the \qty{20}{\kilo\volt} grid to a higher voltage level. \\		
		Load On &		Closing of a load‑switch that connects a specified load to the network. \\		
		Load Off &		Opening of a load‑switch that disconnects the load from the network. \\		
		Capacitor On &		Closing of a switch connecting a shunt capacitor. \\		
		Capacitor Off &		Opening of a switch connecting a shunt capacitor. \\		
		OHL On &		Energization of an OHL line segment. \\		
		OHL Off &		De-Energization of an OHL line segment. \\		
		Cable On &		Energization of a connected cable. \\
		Cable Off &		De-Energization of a connected cable. \\
		Motor Start &		Starting of an induction motor. \\
		\addlinespace[0.1cm]
		1PHG SHC &		Single-phase-to-ground short-circuit. \\
		2PH SHC &		Two-phase (line‑line) short-circuit. \\
		2PHG SHC &		Two-phase-to-ground short-circuit. \\
		3PH SHC &		Three-phase short-circuit. \\
		1PHG HIF &		Single-phase high-impedance ground fault (HIF).\\
		Incipient Fault & Single-phase incipient fault, characterized by high-impedances and self-extinguishing behavior. \\ 
		\bottomrule
	\end{tabular}
\end{table}
\begin{table}
	\centering
	\caption{Parameters used for Event Simulation}
	\label{tab:numeric-params}
	\begin{tabular}{lcc}
		\toprule
		\textbf{Parameter} & \textbf{Typical range (unit)} & \textbf{Source} \\
		\midrule
		
		Load apparent power                & \qtyrange{0.05}{2.5}{\mega\voltampere}                & \cite{BenchmarkSystemsNetwork2014} \\
		Load power factor ($\cos\phi$)                & \qtyrange{0.80}{0.99}{}                 & \cite{BenchmarkSystemsNetwork2014} \\
		Capacitor power rating                       & \qtyrange{0.1}{2.0}{\mega\voltampere}  & \\
		\addlinespace[0.1cm]
		
		OHL conductor spacing                         & \qtyrange{0.4}{2.0}{\meter}             & \cite{flosdorffElektrischeEnergieverteilung2000,BenchmarkSystemsNetwork2014} \\
		OHL tower height                     & \qtyrange{8}{12}{\meter}                & \cite{flosdorffElektrischeEnergieverteilung2000, BenchmarkSystemsNetwork2014} \\
		OHL length                          & \qtyrange{1}{25}{\kilo\meter}           &  \\
		Cable length                      & \qtyrange{0.5}{10}{\kilo\meter}         &  \\
		\addlinespace[0.1cm]
		
		Transformer sat. reactance     & \qtyrange{1.0}{2.0}{\pu}   & \cite{GuidelinesRepresentationNetwork1990} \\
		Transformer sat. exponent      & \qtyrange{7}{18}{\pu}         & \cite{digsilentgmbhPowerFactoryTechnicalReference2024} \\
		Transformer knee flux        & \qtyrange{1.05}{1.25}{\pu}       & \cite{digsilentgmbhPowerFactoryTechnicalReference2024} \\
		\addlinespace[0.1cm]
		
		Petersen-coil tuning                    & \qtyrange{-10}{10}{\ampere}             & \\
		Direct-grounding resistance                      & \qtyrange{0.1}{20}{\ohm}                & \\
		
		\addlinespace[0.1cm]
		
		Arc time constant ($\tau$)                    & \qtyrange{0.2}{0.4}{\milli\second}      & \cite{zhangMulticycleIncipientFault2017} \\
		Char. arc voltage ($u_0$)                 & \qtyrange{300}{4000}{\volt}             & \cite{zhangMulticycleIncipientFault2017} \\
		Char.  arc resistance ($r_0$)              & \qtyrange{0.010}{0.015}{\ohm}           & \cite{zhangMulticycleIncipientFault2017} \\
		\addlinespace[0.1cm]

		Short-circuit resistance                      & \qtyrange{0.001}{20}{\ohm}              & \\
		HIF resistance               & \qtyrange{20}{150000}{\ohm}             & \cite{elkalashyModelingExperimentalVerification2007} \\
		Incipient fault duration                                & \qtyrange{0.002}{0.08}{\second}         & \cite{jannatiIncipientFaultsMonitoring2019} \\
		\bottomrule
	\end{tabular}
\end{table}
To prevent overfitting to specific grids or events, we sample relevant parameters from uniform distributions. The selected ranges for relevant parameters are listed in Tab.~\ref{tab:numeric-params}. Additionally, OHL conductors are randomly sampled from 12 types mentioned by Heuck et al.~\cite{heuckElektrischeEnergieversorgungErzeugung2013} and cables are randomly chosen from PowerFactory's 275 predefined \qty{20}{\kilo\volt} cable types.

Parameters ranges for which no source could be found in literature are estimated. This approach is acceptable because slightly unrealistic parameters do not adversely affect the surrogate task, as feature selection depends on relative signal characteristics rather than absolute parameter accuracy. 

Each parameter configuration is verified by executing a load flow. Non-convergent cases are discarded.

\subsection{Data Generation Process}
We run 20000 electromagnetic transient simulations (EMT) using PowerFactory's Python API. Each simulation runs for \qty{600}{\milli\second} with a step size of \qty{50}{\micro\second}. The initial \qty{100}{\milli\second} allows settling of initial transients, the final \qty{500}{\milli\second} are recorded. The recorded measurements are exported at a standard protection relay sampling rate of \qty{4}{\kilo\hertz} at two locations, namely at the beginning of OHL1a and Cable1a. These positions capture characteristics of transient events both for cable dominated feeders and OHL-based feeders. Additional measurement locations would likely result in redundant data.

Each simulation result is labeled with the event type and the direction as downstream or upstream of the relay for classification in the surrogate task. Therefore, there are 34 distinct event classes in total.

\section{Feature Selection}
\label{sec:feat_selec}
We utilize the library scikit-learn~\cite{pedregosa2011scikit} for all subsequent experiments. Important hyperparameters are listed in Tab.~\ref{tab:hyperfs} and Tab.~\ref{tab:hypercorr} in the Appendix.

\subsection{Considered Features}

\begin{table}[t]
	\centering
	\caption{List of Features}
	\label{tab:features}
	\begin{tabular}{
			p{0.5\columnwidth}
			c
			c
			c
		}
		\toprule
		\textbf{Description} & \textbf{N Feat. x } & \textbf{N Agg.} & \textbf{N Tot.}\\
		 & \textbf{N Chann.} &  &\\
		
		\midrule
		Amplitudes of common harmonic components (1\textsuperscript{st}, 2\textsuperscript{nd}, 3\textsuperscript{rd}, 4\textsuperscript{th}, 5\textsuperscript{th}, 7\textsuperscript{th}, 11\textsuperscript{th}, and 13\textsuperscript{th}) obtained by an FFT. & 8x8 & 6 & 384 \\
		Total harmonic distortion (THD)~\cite{zhangMulticycleIncipientFault2017}. & 1x8 & 6 & 48 \\
		Phase angle difference with respect to $U_a$~\cite{baloujiDistributionNetworkFault2023}. & 1x8 & 6 & 48 \\
		
		\addlinespace[0.15cm]
		Maximum, minimum, mean, standard deviation. & 4x8 & 6 & 192 \\
		Skewness, kurtosis, crest factor, form factor~\cite{wangMultifeatureBasedExtreme2024}. & 4x8 & 6 & 192 \\
		Largest change between consecutive values. & 1x8 & 6 & 48\\
		
		\addlinespace[0.15cm]
		Root mean square of the signal over one period~\cite{baloujiDistributionNetworkFault2023}. & 1x8 & 6 &  48 \\
		Magnitude of symmetric components of the signal ($U_{(0)}, U_{(1)}, U_{(2)}, I_{(0)}, I_{(1)}, I_{(2)}$). & 1x6 & 6 & 36 \\
		Resistance and reactance~\cite{baloujiDistributionNetworkFault2023}. & 2x4 & 6 & 48 \\
		Active and reactive power~\cite{baloujiDistributionNetworkFault2023}. & 2x4 & 6 & 48 \\
		
		\addlinespace[0.15cm]
		Autocorrelation, binned entropy, Fourier entropy, outlier ratio~\cite{christTimeSeriesFeatuRe2018}. & 4x8 &  & 32\\
		
		\addlinespace[0.15cm]
		Detail and approximate coefficients of SWT~\cite{luCableIncipientFault2022}. & 9x8 & 6 & 432 \\
		
		\midrule
		& & \textbf{Total} & 1556 \\
		\bottomrule
	\end{tabular}
\end{table}

As a starting point for the feature selection process, we identified relevant features from literature that are commonly used in the field of incipient fault detection and fault prediction. We implemented a subset of all identified features. Inclusion criteria require sufficient documentation in the respective paper and simplicity in implementation. An example for sufficient documentation is publishing explicit mathematical equations, while simplicity in implementation can be achieved by using existing open source libraries. Exclusion criteria are the necessity of pre-training (e.g., autoencoder-based features), excessive computational or implementation complexity.

A summary of all selected features can be seen in Tab.~\ref{tab:features}, where the column \textit{N Feat. x N Chann.} gives the number of features times the channels for which the features are calculated while the column \textit{N Agg.} gives the number of aggregation functions applied to each feature. Most of these features are computed for each fundamental period individually and are subsequently aggregated over the \qty{500}{\milli\second} window using standard statistical measures, namely minimum, maximum, mean, standard deviation, skewness, and kurtosis. An exception are the detail and approximation coefficients of the stationary wavelet transform (SWT) features implemented as described in~\cite{luCableIncipientFault2022}. These coefficients are calculated across the entire window and then aggregated using the previously mentioned statistical measures. The second exception are the features autocorrelation, binned entropy, and Fourier entropy, and the outlier ratio, which are computed directly for the complete window utilizing the library \textit{tsfresh}, which was developed for time series feature extraction~\cite{christTimeSeriesFeatuRe2018}. Here, the outlier ratio measures the ratio of values which exceed 110\% of the RMS value of the respective channel. Most features are calculated for eight channels, namely the measured phase currents and voltages as well as the zero-sequence current and voltage. Exceptions are symmetrical component based features, which consist of six channels, as well as impedance and power based features which consist of four channels (one for each phase and one for the zero-sequence system).

The features can be grouped into five categories: Frequency domain features (e.g., magnitude of harmonic components, THD, ...), basic statistic measures (e.g., maximum, minimum, ...), features commonly used in the context of power systems (e.g., RMS, resistance, reactance, ...), SWT based features as well as more advanced features for which we utilized the time series library \textit{tsfresh}~\cite{christTimeSeriesFeatuRe2018}.
In total, this approach results in 1556 potential feature candidates.

\subsection{Validation of Correlation between Surrogate Task and Real-World Fault Prediction Task}
The usefulness of our approach critically depends on whether feature selection using simulation data translates to improved FP on real-world measurements. To validate this, we evaluated 1000 randomly selected feature subsets on both tasks  independently using an RF classifier with five-fold cross-validation. Each subsets' usefulness was judged by its accuracy on the simulation task, and its F1 score for the real-world FP task due to the class imbalance caused by the infrequent occurrence of faults.
\begin{figure}
	\centering
	\includegraphics[trim={0 5 0 8},clip,width=\linewidth]{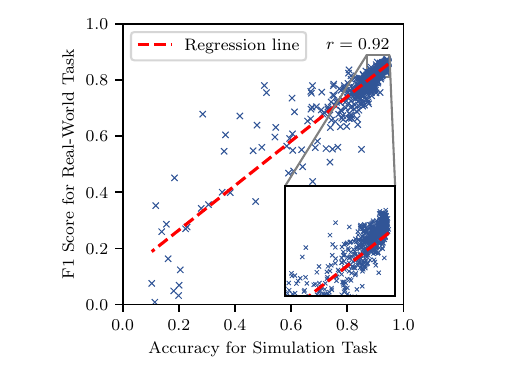}
	\caption{Performance of randomly selected sets of features on the simulation task and the real-world task.}
	\label{fig:correlation}
\end{figure}

The results of this experiment can be seen in Fig.~\ref{fig:correlation}, which demonstrates a strong correlation ($r=0.92$, $p<1e-10$) between feature subset performance on the simulation-based surrogate task and real-world FP effectiveness. This high Pearson correlation coefficient confirms that features optimized for the surrogate task effectively identify precursor symptoms in field measurements. The near-linear relationship demonstrates that simulation-based feature selection accurately predicts real-world utility, eliminating the need for extensive experimentation on scarce field data.
\begin{figure}
	\centering
	\includegraphics[trim={0 5 0 5},clip,width=\linewidth]{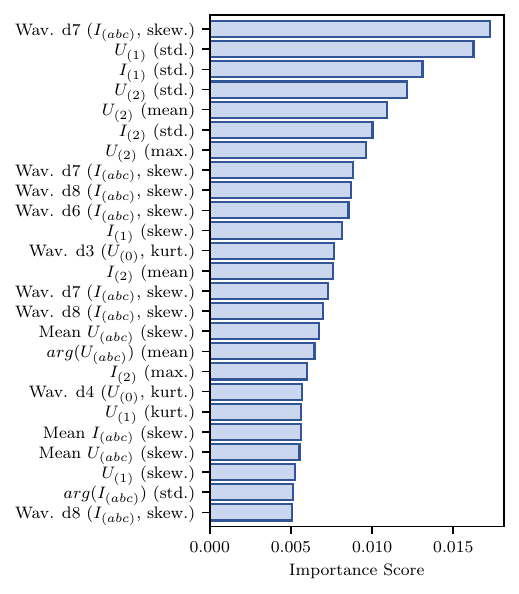}
	\caption{Feature importance scores based on mean impurity decrease of the top 25 features. The signal type (e.g., $I_{abc}$ for one of the phase currents, $U_{(0)}$ for zero-sequence voltage) and the aggregation function are listed where applicable.}
	\label{fig:featureimportances}
\end{figure}

\subsection{Feature Selection Process}
We select a subset of the candidate features based on recursive feature elimination with 5-fold cross-validation (RFE). For this purpose, we utilize scikit-learn's implementation, which is based on the process proposed by Guyon et al.~\cite{guyon2002gene}. During this process, the chosen RF classifier is trained and tested for each of the cross-validation splits. A score, namely the mean accuracy across all five folds is calculated and is recorded. Subsequently, RFE removes the least discriminative $n$ features, determined by the RF's feature importance averaged across all five cross-validation splits. Here, $n$ refers to the step size, which we chose to be one. This process is repeated recursively until no features to remove are left. Finally, the optimal number of features is determined as the one yielding the highest mean accuracy across all five cross-validation splits. Relevant hyperparameters of the feature selection process are listed in the appendix.

\subsection{Selected Features}
\begin{figure}
	\centering
	\includegraphics[trim={0 5 0 5},clip,width=\linewidth]{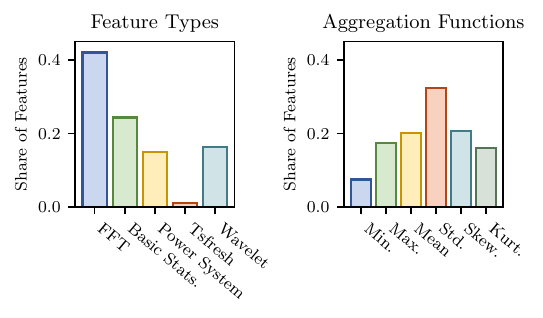}
	\caption{Share of features and aggregation functions in the selected features. The results indicate that standard deviation as an aggregation function and frequency domain features are most useful for fault prediction.}
	\label{fig:featuresummary}
\end{figure}

The feature selection process identified 374 features that optimize performance on the simulation task. Using more or fewer features resulted in degraded classification accuracy. As shown in Fig.~\ref{fig:featuresummary}, FFT-based frequency-domain features represent the largest share of selected features, while standard deviation is the most frequently applied aggregation function. While the feature selection process has no inherent mechanism to establish causation for the selected features, the prevalence of FFT-based features appears consistent with power system theory. Transient events, such as faults with nonlinear arc characteristics or transformer inrush phenomena, induce responses across different frequencies. FFT-based features can capture these higher order harmonics, supporting distinction between relevant and irrelevant events. Similarly, the prevalence of standard deviation-based features is expected, as transient disturbances typically cause deviations from normal operating conditions, making it an effective measure of signal variation.

The impurity-based feature importance scores determined by the trained RF are shown in Fig.~\ref{fig:featureimportances}. Interestingly, six out of the top ten features are symmetrical component based, namely the maximum, mean and standard deviation of the negative sequence voltage, as well as the standard deviation of positive sequence voltage, current and negative sequence current. The remaining top features consist of the skewness of stationary wavelet transform detail coefficients across phase currents. The relatively consistent importance scores across three phase channels for equivalent features (e.g., Wav. d7 ($I_{abc}$, skew.)) validates the approach, as significant disparities would suggest imbalances or artifacts in the dataset.

The prominence of wavelet-based features aligns with previous literature, while symmetrical component based features have received comparatively less attention in related studies. 
\section{Case Study}
\label{sec:casestudy}
The primary objective of the case study is to demonstrate that the previously identified feature set generalizes to FP performance. The utilized dataset, collected with the setup described in Sec.~\ref{sec:data_agg}, contains data from three substations in \qty{22}{\kilo\volt} distribution grids spanning 420 days (Station A), 153 days (Station B), and 116 days (Station C).

\subsection{Training and Test Dataset}
Similar to previous studies, data scarcity is a significant challenge, especially due to the high variability of operational states over a year. Time series cross-validation is unsuitable in this context because most temporal splits would contain data from only one season, potentially leading to models that fail to capture annual operational patterns.  Therefore, we implemented a station-based train-test split strategy that ensures representative seasonal coverage.

Station A, with its continuous year-long recording, captures the full range of seasonal and operational variability and forms the foundation of the training data. Stations B and C, containing shorter recordings, were each split into two equal parts. For training, we combined all data from Station A with the first half of Station B and tested on the second half. The same approach was applied to Station C.

The RF classifier was trained using the optimal feature set containing 374 features. As baselines for comparison, we chose all wavelet-based and all FFT-based features (432 each), because these feature types were used in previous related studies and are comparable in feature count. The pre-processing includes the previously described steps of window selection, feature extraction and aggregation of the previously selected features. Additionally, data from each station are standardized individually by subtracting the mean and dividing by the standard deviation for each feature.

\subsection{Evaluation}
Faults are considered as distinct events if they are separated by at least \qty{24}{\hour}. A fault is counted as correctly predicted (true positive) if the classifier outputs a probability above 0.5 for a \qty{5}{\hour} moving average in the 3.5 days before a fault. Although the full observation window spans seven days, predictions are only attributed to a fault if they fall within this shorter interval to ensure their relevance.  The prediction is counted as false positive if the classifier predicts a probability above 0.5 but no fault materializes within the subsequent ten days, to consider the variability of the underlying physical phenomena.
\begin{figure}
	\centering
	\includegraphics[width=1\linewidth]{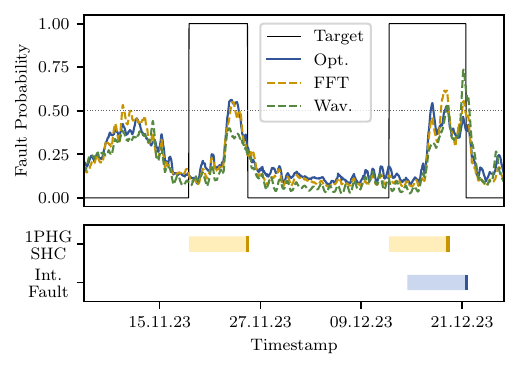}
	\caption{Two exemplary faults and the output of random forest classifiers for the optimal feature set (blue), FFT-based features (orange) and wavelet-based features (green).}
	\label{fig:faultpredexample}
\end{figure}

\subsection{Results}
Tab.~\ref{tab:result_metrics} shows that the optimal feature set selected during feature selection outperforms the two baselines both on each station individually and combined. While it performs better on Station B, the results on Station C are an improvement over the baseline and in an acceptable range. 

\begin{table}[b]
	\centering
	\caption{Performance Metrics by Feature Set, Combined and Individually for Station B and C in Brackets}
	\begin{tabular}{lccc}
		\toprule
		Feature Set & Precision & Recall & F1 Score \\
		\midrule
		Opt. Feat.      & 0.80 (1.00/0.67) & 0.80 (0.80/0.80) & 0.80 (0.89/0.73) \\
		FFT Feat.       & 0.75 (0.67/1.00) & 0.60 (0.80/0.40) & 0.67 (0.73/0.57) \\
		Wavelet Feat.   & 0.71 (1.00/0.60) & 0.50 (0.40/0.60) & 0.59 (0.57/0.60) \\
		\bottomrule
	\end{tabular}
	\label{tab:result_metrics}
\end{table}

Fig.~\ref{fig:faultpredexample} visualizes two representative fault events. Wavelet-based features result in fault prediction performance with high variance. FFT-based features cause false positive predictions. In contrast, the optimal feature set is relatively robust. An interesting pattern observed is a decrease in fault probability shortly before the actual fault occurrence. This pattern may be caused by the \qty{5}{\hour} moving average smoothing effect and the labeling convention, which only labels the precursor symptoms as positive, but not the fault event itself. Consequently, features measured shortly before faults may be more similar to the actual fault event, which is labeled as negative, than precursor symptoms observed up to seven days prior.

\begin{figure}
	\centering
	\includegraphics[width=1\linewidth]{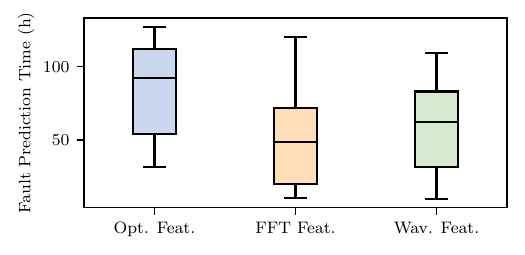}
	\caption{Distribution of fault detection lead times in hours across Stations A, B, C, and the combined dataset.}
	\label{fig:faultpredictionboxplot}
\end{figure}

As the practical usefulness of FP depends on how far in advance a fault can be anticipated, the distribution of hours between the first positive prediction and the actual fault is visualized in Fig.~\ref{fig:faultpredictionboxplot}. The optimal feature set predicts faults with a mean lead time of \qty{84.8}{\hour} compared to \qty{54.4}{\hour} and \qty{59.2}{\hour} for FFT-based and wavelet-based features respectively.

\subsection{Impact of Classifier and Prediction Horizon}
\begin{figure}
	\centering
	\includegraphics[width=\linewidth]{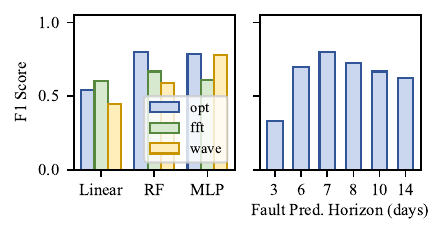}
	\caption{Combined F1-Score of Station B and C for a logistic regression (Linear), random forest (RF), and multi-layer-perception (MLP) classifier as well as for different prediction horizons utilizing the chosen RF classifier and the optimal feature set.}
	\label{fig:combinedplot}
\end{figure}

Both the classifier and the prediction horizon were chosen based on experiments from previous literature. While not the focus of this paper, we conducted a small number of initial experiments to analyze the sensitivity of our results to these choices by repeating the previously mentioned case study with different classifiers and fault prediction horizons. The achieved F1-scores aggregated over Station B and C for different models and fault prediction horizons are presented in Fig.~\ref{fig:combinedplot}.

We chose a significantly simpler, linear model, namely a logistic regression fitted with stochastic gradient descent, and a more complex model, namely a multi-layer-perceptron (MLP). The results indicate that the simpler linear model cannot predict faults as effectively, while the MLP achieves similar results as the RF classifier. Although feature selection depends on the classifier, the features we selected appear relatively robust, while FFT‑ and wavelet‑based features vary more across classifiers.

The F1 score drops significantly if the fault prediction horizon is too short but is less sensitive to longer fault prediction horizons. Overall, our initial experiments indicate agreement with previous literature concerning the chosen classifier and a prediction horizon of seven days. Nevertheless, more comprehensive and rigorous experiments are necessary for a definitive conclusion.

\section{Discussion}
\label{sec:discussion}
\subsection{Limitations}
While the case study demonstrates the effectiveness of the selected features, several limitations must be acknowledged. First, the classifier and the prediction horizon were chosen based on related studies and are not optimized in this paper.

Second, feature selection was performed on a surrogate simulation task, a necessary compromise to mitigate overfitting risks on scarce field data. The strong correlation (r = 0.92) between simulation task performance and real-world FP effectiveness validates this approach, though the imperfect correlation indicates that simulation-optimized features may not fully capture all nuances of real-world precursor symptoms. The strong correlation is likely to hold for different fault types and operating conditions, as the real-world dataset contains different fault types as well as data from different seasons. However, as real-world experiments are currently limited to three substations, there is a risk that real-world performance on different substations lacks behind the performance on simulation data. This risk increases if new substations differ significantly from the tested ones, for example when transitioning from resonance grounded to effectively grounded grids. When abundant real-world data becomes available, direct feature selection on field data can further improve FP performance.

Scarce real-world fault data from three substations may also limit the case study's robustness. While the station-based train-test split ensures seasonal coverage while avoiding data-leakage from train to test dataset, results may vary when additional substations are incorporated, potentially leading to improved or diminished performance. Future work will incorporate more data to assess the robustness of the findings.

Additionally, not all faults exhibit detectable precursor symptoms, yet the current labeling approach treats all faults as predictable to eliminate manual labeling requirements. This limitation should be addressed by a labeling scheme that distinguishes between predictable and unpredictable faults.

Finally, the FP pipeline currently only utilizes voltage and current measurements. It is likely that additional data sources like dissolved gas analysis for transformers, acoustic fingerprints or weather conditions could improve the predictive power of the classifiers. The integration of additional data sources should therefore be considered in further studies.

Despite these limitations, the relative advantage of the selected features over the baselines demonstrates the validity of the simulation-based feature selection methodology.

\section{Conclusion} 
\label{sec:conclusion}
This study addressed the challenges of data scarcity and a lack of systematic feature comparison for FP. We implemented a simulation-based feature selection and demonstrated its effective transfer to real-world applications using field data from three substations. We identified 374 optimal features from an initial set of 1556 candidates, with wavelet-based and symmetrical component-based features proving most important for prediction performance. The successful simulation-to-real transfer indicates that insights from previous incipient fault detection studies, which mainly rely on simulation data, can be transferred to fault prediction approaches. 

A mean prediction lead time of \qty{84.8}{\hour} was achieved in this study, meaning a fault can be predicted approximately 3.5 days in advance. This time span can be utilized to address issues in the grid before they lead to high current faults. This proactive fault management can reduce outage duration and improve grid reliability. The current development of IEC 61850-compliant smart grid infrastructure allows a simple integration of fault prediction approaches into existing hardware.

\section*{Acknowledgment}
This project was partially funded by the Deutsche Forschungsgemeinschaft (DFG, German Research Foundation) - Project number 535389056.

\renewcommand{\thetable}{A\arabic{table}}

\section*{Appendix: Hyperparameters}
\label{sec:appendix}
\setcounter{table}{0}

Tab.~\ref{tab:hyperfs} and Tab.~\ref{tab:hypercorr} list the hyperparameters for the classifiers, cross-validation, feature selection algorithm and evaluation metrics. For hyperparameters not listed in the tables, scikit-learn default values were used.
\begin{table}[ht]
	\centering
	\caption{Hyperparameters used for feature selection}
	\label{tab:feature_selection}
	\begin{tabular}{lll}
		\toprule
		\textbf{Component} & \textbf{Hyperparameter} & \textbf{Value} \\
		\midrule
		RF Classifer & \texttt{n\_estimators} & 100 \\
		& \texttt{random\_state} & 42 \\
		& \texttt{n\_jobs} & $-1$ \\
		\addlinespace
		Cross-Validation & \texttt{Type} & Stratified K-Fold \\
		for RFE & \texttt{n\_splits} & 5 \\
		& \texttt{shuffle} & True \\
		& \texttt{random\_state} & 42 \\
		\addlinespace
		RFE & \texttt{step} & 1 \\
		& \texttt{cv} & Defined above \\
		& \texttt{n\_jobs} & $-1$ \\
		& \texttt{scoring} & Accuracy \\
		\bottomrule
	\end{tabular}
	\label{tab:hyperfs}
\end{table}

\begin{table}[ht]
	\centering
	\caption{Hyperparameters for correlation analysis between simulation task and real-world task on data from Station A}
	\label{tab:correlation_stationA}
	\begin{tabular}{lll}
		\toprule
		\textbf{Component} & \textbf{Hyperparameter} & \textbf{Value} \\
		\midrule
		RF Classifier for  & \texttt{n\_estimators} & 100 \\
		Simulation and & \texttt{random\_state} & 42 \\
		Real-World Task & \texttt{n\_jobs} & $-1$ \\
		\addlinespace
		
		Cross-Validation for & \texttt{Type} & Stratified K-Fold \\
		Simulation Task & \texttt{n\_splits} & 5 \\
		& \texttt{shuffle} & True \\
		& \texttt{random\_state} & 42 \\
		\addlinespace
		Cross-Validation for & \texttt{Type} & Time-Series Split \\
		Real-World Task & \texttt{n\_splits} & 5 \\
		& \texttt{shuffle} & False \\
		\addlinespace
		Evaluation Metrics & \texttt{scoring\_sim} & Accuracy \\
		& \texttt{scoring\_real} & F1-Score \\
		\bottomrule
	\end{tabular}
	\label{tab:hypercorr}
\end{table}


\end{document}